\documentclass[12pt]{article}
\usepackage{amssymb}
\usepackage[dvips]{graphics}
\begin{document}
\def\theequation{\arabic{section}.\arabic{equation}} 
\hyphenation{ge-ne-ral}
\hyphenation{as-so-cia-ted}
\hyphenation{gua-ran-tee}
\def\theequation{\arabic{section}.\arabic{equation}}
\newcommand{\be}{\begin{equation}}
\newcommand{\ee}{\end{equation}}
\begin{titlepage}
\setcounter{page}{1}
\title{Phantom cosmology with general potentials}
\author{Valerio Faraoni \\ \\
{\small \it $^1$ Physics Department, Bishop's University}\\
{\small \it Lennoxville, Qu\`{e}bec, Canada J1M~1Z7}\\
{\small vfaraoni@cs-linux.ubishops.ca }
}
\date{}
\maketitle   %\thispagestyle{empty}  \vspace*{1truecm}
\begin{abstract}
We present a unified treatment of  the phase space of a 
spatially flat homogeneous and isotropic universe dominated by a phantom 
field. Results on the dynamics and the late time attractors  (Big Rip, 
de Sitter, etc.) are derived  without specifying the form of the phantom 
potential, using only general assumptions on its shape. Many results found 
in the literature are quickly recovered and predictions are made for new 
scenarios.
\end{abstract}
\vspace*{1truecm} \begin{center}  
%{\bf Keywords:}
PACS numbers: 98.80.-k, 04.90.+e
% cosmology, other topics in GR
\end{center}     
\end{titlepage}  
 \clearpage 
\setcounter{page}{2}

\section{Introduction}
\setcounter{equation}{0}

There is now substantial agreement that the expansion of the universe is 
accelerated, with evidence from type Ia 
supernovae \cite{SN}, {\em WMAP} data \cite{WMAP}, and large scale 
structure surveys \cite{LSS}.
Two classes of models aim at explaining the observed cosmic 
acceleration: one modifies gravity on large scales by introducing 
corrections to the Einstein-Hilbert \cite{modifiedgravity} or  
the Palatini \cite{Palatini} action (often there are instabilities  
\cite{mginstabilities} or  incorrect  post--Newtonian limits 
\cite{mglimit}), and 
the other advocates the 
existence of dark energy with density $\rho$ and exotic pressure 
$P<-\rho/3$.

Most dark energy candidates proposed thus far are scalar fields; there is 
marginal evidence \cite{Melchiorrietal03} for an 
effective equation of state parameter of the dark energy $w \equiv P/\rho 
<-1$, which is equivalent to increasing Hubble parameter $\dot{H}>0$. 
If confirmed, this measurement is important  because an 
increasing $H$ cannot be explained by Einstein gravity with a canonical 
scalar field \cite{Caldwell,superq,mybook}. If the 
universe really superaccelerates it may end its 
existence in a finite time in an explosive expansion of the 
scale factor accompanied by diverging dark energy density, called a Big 
Rip \cite{Caldwelletal,Carrolletal}. Other kinds 
of singularities at a finite time in the future  discussed in the 
literature  are called ``sudden future singularities'' and are 
characterized by finite scale factor and  Hubble parameter,  
and diverging $\dot{H}$ and dark  
pressure \cite{Barrow04}.

Simple models of a superaccelerated universe employ a  
scalar field coupled nonminimally to the 
Ricci curvature  
\cite{superq,FaraoniPRD2003,mybook,FujiiMaeda,CarvalhoSaa} or 
a {\em phantom} field, i.e., a minimally coupled scalar field with  
negative kinetic energy \cite{Caldwell,Caldwelletal,Carrolletal}. 
Phantom  fields and fields with non--canonical energy are present in 
string theories
\cite{MersiniBasteroKanti,BasteroFramptonMersini,Frampton03} and 
supergravity \cite{Nilles84}, are 
associated to bulk 
viscosity due to particle production \cite{Barrow88}, or arise in higher 
order theories \cite{Pollock88}.  It has also been proposed that early 
inflation and late time acceleration of the universe can be unified in a 
single theory based on a phantom field \cite{Sergei03}.
A phantom field poses several 
challenges: 
it is subject to severe instabilities, which may  
perhaps be avoided by thinking of the phantom as an effective field theory 
resulting from some fundamental theory with positive energies 
\cite{Sergei03}. 
A fundamental quantum phantom is very difficult to stabilize 
\cite{Carrolletal,Clineetal}.

In general, a form of dark energy with $w<-1$ is worrisome because it 
violates the energy conditions cherished by most physicists and opens the 
door 
to disturbing exotica. For example, even a small amount of exotic matter 
violating the weak energy condition may be sufficient to open up a 
wormhole and make time travel possible \cite{VisserKarDadhich03}. On the 
other hand, even a simple classical scalar field coupled nonminimally to 
the Ricci curvature may violate all of the energy conditions 
\cite{BarceloVisser}, and one should probably not be too rigid in 
rejecting phantom fields {\em a priori}.

The peculiar dynamics arising from a negative kinetic energy density has 
been studied with the help of a toy model consisting of two 
mutually coupled 
oscillators, one with negative kinetic energy representing the phantom 
field, and the other with positive kinetic energy representing the 
gravitational 
field \cite{Carrolletal}. However, this toy model fails
to properly describe the dynamics \cite{oscillators} and it is 
preferable to study the actual dynamical system. Furthermore,  
a number of papers in the literature study phantom models 
with different specific potentials \cite{phantommodels}. Here we study the 
dynamics without toy 
models or approximations  and we deduce the behaviour and the late time 
state of a phantom--dominated universe {\em without specifying the form of 
the potential}. 
The assumption of a spatially flat 
Friedmann--Lemaitre--Robertson--Walker (FLRW) universe  plus general 
assumptions on 
the potential (such as boundedness, monotonicity, etc.)  
allow one to derive the asymptotic dynamics. In addition to the need 
to understand phantom dynamics with 
general potentials, our study is motivated by the question whether the 
universe will end 
in a Big Rip or will expand forever. In fact, a universe that 
superaccelerates  may end its existence in a Big Rip or 
it may expand forever, depending on the dynamical equations. 
Often authors working with a specific choice of the phantom potential 
find a de Sitter regime as the final stage of evolution (late time 
attractor).  We show that certain features of the 
phantom potential determine whether the final state is an 
indefinite expansion  or a Big Rip, and the analysis needs not be 
repeated for all the possible potentials.  We adopt the notations and 
conventions of \cite{Wald} and units such  that $c=G=\hbar=1$.

\section{ The phase space of phantom cosmology}
\setcounter{equation}{0}

Based on the recent cosmological observations we adopt 
the FLRW line element 
\be \label{1}
ds^2=-dt^2 +a^2(t) \left( dx^2 + dy^2 +dz^2 \right) 
\ee
in comoving coordinates $\left( t,x,y,z \right)$. We consider the 
situation in which dark energy has already come to dominate the cosmic 
dynamics, hence a phantom field is the only form of matter 
in the Einstein equations. The energy density and pressure of the 
phantom are, respectively,
 \be \label{2}
\rho =-\, \frac{1}{2} \, \dot{\phi}^2 +V( \phi) \;, \;\;\;\;\;\;\;
P =-\, \frac{1}{2} \, \dot{\phi}^2 - V( \phi) \;.
\ee
The distinguishing  negative 
kinetic energy  is evident in eqs.~(\ref{2})  which exhibit the 
``wrong'' sign for the kinetic term $\dot{\phi}^2/2$.
The Einstein equations for $\phi(t)$ and the scale factor $a(t)$ 
are  
\begin{eqnarray}
&& H^2  = \frac{\kappa}{6} \left( -\dot{\phi}^2 +2V \right) \;, \label{4} 
\\
&& \nonumber \\
&& \dot{H}+H^2=\frac{\kappa}{3} \left( \dot{\phi}^2 +V \right) \;, 
\label{5} \\
&& \nonumber \\
&& \ddot{\phi}+3H\dot{\phi}-\, \frac{dV}{d\phi}=0 \;, \label{6}
\end{eqnarray}
where $H \equiv \dot{a}/a $ is the Hubble parameter, an overdot denotes 
differentiation with respect to the comoving time $t$, and $\kappa \equiv 
8\pi G$. The sign of the term $dV/d\phi$ in the Klein--Gordon equation 
(\ref{6}) is the opposite of the usual one. The potential is required to 
be positive 
by eq.~(\ref{4}), and this modification of the usual Klein--Gordon 
equation implies 
that a phantom falls up in an increasing potential, is repelled by  
a minimum, and settles in a maximum of $V$  
\cite{Caldwelletal,SinghSamiDadhich03,SamiToporensky03}. In addition, a 
monotonically increasing unbounded potential may be expected to generate 
runaway 
solutions -- the analogy with a canonical scalar field in a negative 
potential not bounded from below applies (these properties are proven 
later in this paper).

Eq.~(\ref{4}) implies that the phantom energy density is non--negative for 
any solution of eqs. (\ref{4})--(\ref{6}). Only two equations in the set 
(\ref{4})--(\ref{6}) are independent --  when $\dot{\phi} \neq 0$ 
one can derive the Klein--Gordon equation (\ref{6}) from the other two.

The field equations can be obtained from the Lagrangian $L$ or the 
Hamiltonian ${\cal H}$
\be \label{7}
L=3a\dot{a}^2 +\kappa \, a^3 \left[ \frac{ \dot{\phi}^2}{2} +V( \phi) 
\right] \;, \;\;\;\;\;
{\cal H}= a^3 \left[ H^2 +\frac{\kappa}{6} \, \dot{\phi}^2 -\frac{\kappa 
V(\phi)}{3} \right] \;,
\ee
the Hamiltonian constraint (\ref{4}) corresponding to ${\cal H}=0$. 
The orbits  of the solutions of eqs.~(\ref{4})--(\ref{6}) are constrained 
to the surface of constant energy ${\cal H}=0$ in the phase space 
$\left( a, \dot{a}, \phi , \dot{\phi} \right)$. We choose as dynamical 
variables the Hubble parameter and the scalar field $\left( H, \phi 
\right)$ which correspond to physical observables\footnote{Other authors 
choose as 
dynamical variables various combinations of $H$ and $\phi$ but this 
obscures the physical interpretation of their results.}. The phase space 
accessible to the orbits of the solutions is the 
two-dimensional energy surface $ {\cal H}=0  $  in  the three-dimensional 
space $ \left( H, \phi, \dot{\phi} \right)$; this surface is in general 
curved (the appendix provides an example).

Once the values of $H$ and $\phi$ are given, one computes the 
corresponding value(s) 
of $\dot{\phi}$ by rewriting the Hamiltonian constraint 
(\ref{4}) as the algebraic equation for $\dot{\phi}$ 
\be 
\dot{\phi}^2 -2V +\frac{6H^2}{\kappa}=0 \;,
\ee
with  solutions
\be \label{9}
\dot{\phi}_{\pm}\left( H, \phi \right)=\pm \sqrt{ 2\left[ 
V(\phi)-\frac{3H^2}{\kappa} \right] } \equiv \pm \sqrt{F \left( H, \phi 
\right)} \;.
\ee
Depending the form of $V \left( \phi \right) $, there may be a region 
${\cal F}$  of the  phase space forbidden to the dynamics and 
corresponding to a negative sign of $ F \equiv 2\left( V-3H^2/\kappa 
\right)$:
\be
{\cal F} \equiv \left\{ \left( H, \phi, \dot{\phi} \right): \;\;\;\;\;
V(\phi)< \frac{3H^2}{\kappa} \; \right\} \;.
\ee
When $F \left( H, \phi \right) >0$ there are two distinct values of 
$\dot{\phi}$ for each value of $H$ and $\phi$, corresponding to the 
fact that the two-dimensional phase space consists of two sheets 
joining each other only at the points of the set
\be 
{\cal B} \equiv \left\{ \left( H, \phi, \dot{\phi} \right): \;\;\;\;\;
V(\phi) =  \frac{3H^2}{\kappa} \;, \;  \dot{\phi}=0 \;\right\} \;,
\ee
which constitutes the boundary of the forbidden region ${\cal F}$ and 
lies in the $\left( H, \phi \right)$ plane.
We denote the sheet corresponding to the positive sign of $\dot{\phi}$ 
``upper sheet'', while the other is called ``lower sheet''. This structure 
of the phase space is general in scalar--tensor cosmology 
\cite{STphasespace} and is well known in 
the theory of a scalar field with 
canonical kinetic energy coupled non--minimally to the Ricci curvature 
 \cite{Amendolaetal,Foster,Gunzigetal}.
Phantom cosmology can be seen as a special case of  scalar--tensor theory 
for which there is always a forbidden region in the phase 
space, while this needs not be the case in other scalar--tensor theories. 
The equilibrium points of the system (\ref{5})--(\ref{6}) consist of de 
Sitter spaces with constant scalar field $ \left( H_0, \phi_0 \right)$. If 
they exist, these fixed points 
satisfy the constraints imposed by eqs.~(\ref{4})--(\ref{6})
\be \label{12}
H_0  = \pm \sqrt{ \frac{\kappa V_0}{3}} \;,\;\;\;\;\;\;
V_0' =  0 \;,
\ee 
where $V_0 \equiv V( \phi_0) $ and $V_0'\equiv \left.  \frac{dV}{d\phi} 
\right|_{\phi_0}$. The Hamiltonian constraint  (\ref{4}) can only be 
satisfied if $V( \phi) \geq 0$, which guarantees that $H_0$ is real -- 
then there are de Sitter 
fixed points provided that $dV/d\phi$ has zeros (the existence of 
equilibrium points is not guaranteed in all scalar--tensor 
theories). These fixed points lie on the boundary ${\cal B}$ of the 
forbidden region\footnote{In other scalar--tensor theories the fixed 
points, if they exist, may be located anywhere on the energy surface 
${\cal H}=0$.}.

What happens to an orbit with initial conditions chosen 
exactly on ${\cal B}$~? To answer this question we consider the tangent to 
the orbits
\be \label{tangent}
\vec{T} =\left( \dot{H}, \dot{\phi}, \ddot{\phi} \right)=\left( 
\frac{\kappa}{2}\,\dot{\phi}^2, \dot{\phi}, \frac{dV}{d\phi}-3H\dot{\phi} 
\right) 
\ee
in the  $\left( H, \phi, \dot{\phi} \right)$ space. On ${\cal B}$ it is 
$\dot{\phi}=0$ and $\vec{T}_{{\cal B}}=\left( 0,0, dV/d\phi \right)$. 
Hence if 
$\left.  dV/d\phi \right|_{ {\cal B}}>0$, an orbit beginning on ${\cal 
B}$ will move into the upper sheet, while if 
$\left.  dV/d\phi \right|_{ {\cal B}} < 0 $, the orbit will move into the 
lower sheet. If instead $ \left. 
dV/d\phi \right|_{ {\cal B}} = 0$ one 
has a fixed point satisfying the properties $\dot{H}=0$, $\dot{\phi}=0$, 
$H^2=\kappa V/3$, and 
$V'=0$.

The following result is true for any form of the potential $V( \phi 
) $:  $H$ {\em always increases along the orbit of 
any solution except at points where $\dot{\phi}=0$, at which also 
$\dot{H}$ vanishes.}

This follows by combining eqs.~(\ref{4}) and (\ref{5}) to  
obtain\footnote{Eq.~(\ref{15}) is equivalent to eq.~(7) of 
\cite{GuoPiaoZhang} multiplied by $\dot{\phi}$.}
\be  \label{15}
\dot{H}= \frac{\kappa}{2} \, \dot{\phi}^2 \;,
\ee
hence $\dot{H}>0$ except at the points where $\dot{\phi}=0$, in 
particular the fixed points $ \left(H_0, \phi_0 \right)$.

Eq.~(\ref{15})  is expressed by saying that the universe always {\em 
superaccelerates} (i.e., $\dot{H}>0$ as opposed to acceleration defined by 
$\ddot{a}=\dot{H}+H^2 >0$), or that the phantom field is a form of 
{\em superquintessence}  \cite{superq,mybook}.   Eqs.~(\ref{4}) and 
(\ref{15}) imply that 
\be \label{16}
\dot{H}+3H^2=\kappa \, V \;.
\ee

A second result is that {\em there are no limit cycles (periodic 
orbits).}

In fact, $H(t)$ is non--decreasing and hence it cannot 
periodically return to its initial value apart from the trivial case of a 
fixed point. If $dV/d\phi$ has definite sign this possibility  
is also excluded, cf. eq.~(\ref{12}).

\section{Bounded potential}
\setcounter{equation}{0}

{\em If the potential 
$V(\phi)$ is bounded from above by 
a (positive) constant $V_0$, then the asymptotic solution of 
eqs.~(\ref{4})--(\ref{6}) at large times is such that $\dot{H} \rightarrow 
0$ and $\dot{\phi}\rightarrow 0$. The resulting de Sitter attractor 
$\left( H_0, \phi_0 \right)$ is a global attractor.}

In fact, according to eq.~(\ref{16}), $\dot{H}=\kappa V -3H^2$; 
assume that $\dot{H}$ does not tend to zero (or that it does not 
tend to zero faster than $t^{-1}$) as 
$t\rightarrow + \infty$. Then the limit
\be
\lim_{t\rightarrow +\infty} H(t)= \lim_{t\rightarrow 
+\infty } \int_{t_0}^{t} \dot{H}( \tau) 
\, d\tau  = + \infty 
\ee
since $\dot{H}>0$. But then $V(t)>3H^2/\kappa \rightarrow +\infty$ as 
$t\rightarrow +\infty$ and this contradicts the fact that $V$ is bounded 
from above. Hence it must be $\dot{H}\rightarrow 0$ (faster than $t^{-1}$)  
as $t \rightarrow +\infty$.

Since $\dot{H}=\kappa\dot{\phi}^2/2$, then $\dot{\phi}\rightarrow0$ 
(faster than $t^{-1/2}$) as well  when $t\rightarrow +\infty$. The 
asymptotic state for a potential bounded from above is one with $\dot{H} 
\rightarrow 0 $ and $\dot{\phi} \rightarrow 0$. Eq.~(\ref{16})  gives the 
asymptotic relation $H_0^2=\kappa V/3$ and the Klein--Gordon equation 
reduces, in this limit, to $ \left. dV/d\phi \right|_{\phi_0}=0$  and the 
asymptotic state is 
the de Sitter fixed point $ \left( H_0, \phi_0 \right)$. The function
\be
L \left( H, \phi \right)=\left( H-H_0 \right)^2 
\ee
defined on the entire phase space is a Ljapunov function -- in fact
$ L( H, \phi )>0  $ $ \forall \left( H,\phi \right) \neq \left(  H_0, 
\phi_0 
\right) $, $ L \left( H_0, \phi_0 \right) =0 $, and
$ dL/dt =\kappa \dot{\phi}^2  \left( H-H_0 \right)
 \leq 0 \; \forall 
\left( H,\phi \right) \neq \left( H_0, \phi_0 \right) $
because $H$ tends to $H_0$ while non--decreasing, hence $H\leq H_0 $ 
$\forall t$. The system is 
stable and the attraction basin of the de Sitter attractor is the entire 
phase space. This excludes the possibility of a Big Rip or a 
sudden future singularity.
\vskip0.3truecm

If the potential has maxima, minima, or inflexion points, there will be 
fixed 
points of the system (\ref{4})--(\ref{6}). The character of these points 
determines the stability of the fixed point. One can assess the stability 
by considering the perturbed equations (\ref{5})--(\ref{6}) for 
homogeneous  
perturbations that depend only on time, as done in 
\cite{HaoLi03PRD2,HaoLiastroph0309746}. It is more 
meaningful to study 
stability against inhomogeneous 
perturbations which depend on both space and time. Then the  complication 
arises that a gauge-independent formalism is required. The  study of
inhomogeneous perturbations  is carried out in 
\cite{deSitter} for generalized gravity theories described by the action
\be 
S=\int d^4x \, \sqrt{-g} \left[ \frac{1}{2} \, f(\phi, R) 
-\frac{\omega(\phi) 
}{2}g^{ab} \nabla_a\phi 
\nabla_b \phi -V( \phi) \right] \;.
\ee
The result of \cite{deSitter} is that there is stability if and 
only if 

\be
\frac{
\frac{ f_{\phi\phi}(\phi_0)}{2} -V_{\phi\phi}(\phi_0) +\frac{6f_{\phi 
R}(\phi_0) 
H_0^2}{f_R( \phi_0)} }{\omega_0 \left( 1+\frac{ 3f_{\phi R}^2( 
\phi_0)}{2\omega_0 f_R(\phi_0)} \right) } \leq 0 \;.
\ee

Phantom cosmology is the special case $f(\phi, R ) =R$, $\omega =-1$, 
$f_{\phi R}=f_{\phi\phi}=0$, $f_R=1$ and stability corresponds to $V_0'' 
\leq 0$, where a prime denotes differentiation with respect to $\phi$. 
 The  perturbation analysis of \cite{deSitter} establishes that the  
de Sitter fixed  point is stable against inhomogeneous linear 
perturbations if $V(\phi)$ has a maximum there. Here we  recover this 
result in a different approach that does not rely on a linear 
perturbation analysis and it is valid to  any order,  extending 
and complementing the results of \cite{deSitter} applied to  phantom 
cosmology. Moreover, \cite{deSitter} does 
not draw conclusions  about the size of the attraction basin, while here 
we establish that this attraction basin is the 
entire phase space. This 
agrees with, and extends,  the result of Guo {\em et al.} 
\cite{GuoPiaoZhang} that, if a late time attractor exists in the phase 
space of phantom cosmology with bounded potential, it is unique.

The  phantom field settling in a maximum and producing a 
late time de Sitter attractor is consistent with the literature selecting 
specific potentials $V(\phi)$ and with 
general conjectures  
\cite{SamiToporensky03,Carrolletal,SinghSamiDadhich03}. Let us 
see some examples.
 Singh, Sami and Dadhich \cite{SinghSamiDadhich03} consider the 
bell-shaped potential $ V(\phi)= V_0 /  \cosh \left( \kappa \alpha 
\phi \right) $ attaining its global maximum  $V_0$ at $\phi=0$, and they 
find a 
late time de Sitter attractor. The same type of attractor is found by 
Carroll,  Hoffman and Trodden \cite{Carrolletal} for the Gaussian 
potential  $V(\phi)=V_0 \, \mbox{e}^{-\phi^2/\sigma}$.

Hao and Li \cite{HaoLiastroph0309746} 
consider the inverse power-law potential $ 
V( \phi) = V_0-\sigma \left( \phi / \phi_0  \right)^2 $ 
for $  \left| \phi \right| \leq \phi_0 \sqrt{ \frac{V_0}{\sigma} } $, with 
maximum $V_0$ at $\phi=0$. They find again a de Sitter attractor 
empty of scalar field ($\phi=0$) and with a truly constant cosmological 
constant $\Lambda = V_0$.

Another situation in which the potential is bounded from above occurs when 
$V( \phi)$ has a horizontal asymptote $V_0$  approached from below 
as $\phi \rightarrow \pm \infty$. Under these conditions the scalar field 
rolls up the slope of the potential but its (negative) kinetic energy is 
dissipated by friction -- described by the term $3H\dot{\phi}$ in 
eq.~(\ref{6}) -- while the force term $-dV/d\phi$ tends to zero 
approaching 
the asymptote, with $3H\dot{\phi} \simeq dV/d\phi$ and $\dot{\phi} 
\rightarrow 0$. As a result the motion stops ($\dot{\phi}\rightarrow 0$) 
and 
a de Sitter regime with constant $\phi$ and $H$ is approached. 
An example is the potential studied by Sami and 
Toporensky \cite{SamiToporensky03} $ 
V(\phi) =V_0 \left[ 1-\mbox{e}^{-c\phi^2} \right] $, 
which has the shape of an inverted bell with  $V_0$ as  
horizontal asymptote. These authors  find  a 
late time de Sitter attractor with cosmological constant $\Lambda= 
V_0$ \cite{SamiToporensky03}.

Guo {\em et al.} \cite{GuoPiaoZhang} consider the bounded potential $ 
V( \phi )=V_0 \left[ 1+\cos \left(  \phi / f  \right) \right] $
and find a slow-climb solution in which $\phi$ settles in the maximum of 
the potential at $\phi=0$ in a de Sitter regime with $\left( H_0, \phi_0 
\right)=\left( \sqrt{ \kappa V_0/3 } , 0 \right)$.

\section{Unbounded potential}
\setcounter{equation}{0}

If the phantom potential is not bounded from above it would seem that the 
asymptotic state could be either a de Sitter regime, or a very 
different state, possibly a Big Rip, 
depending on the shape of $V( \phi)$ and its slope. 
Physically, the motion of $\phi(t)$ is analogous to the motion of  a ball 
with negative kinetic energy falling up a hill under the force 
$dV/d\phi$, damped  by the term $3H\dot{\phi}$. If this  
friction becomes negligible in comparison with the force and inertial  
terms, 
$3H\dot{\phi} << 
dV/d\phi , \ddot{\phi} $, one has a ``slow-climb'' regime analogous to the 
slow-roll, potential-dominated, regime of inflation for a canonical scalar 
field \cite{SamiToporensky03,GuoPiaoZhang}. This happens if the 
potential becomes sufficiently steep. If instead friction is 
comparable to the 
force term, motion 
could stop ($\dot{H} \sim 0 $, $\dot{\phi} \sim 0$)  
corresponding to a de Sitter regime. If $V$ is strictly increasing  this 
situation is forbidden and the expansion is 
super--exponential: {\em if $V(\phi)$ is not bounded from above and is 
strictly increasing the solution of eqs.~(\ref{4})--(\ref{6}) cannot 
approach a de Sitter fixed point at late times, but $H$ and $\phi$ go to 
infinity.}

In fact, a de Sitter fixed point has $ V'=0 $ and in 
order to approach a de Sitter fixed point  it must be $V' \rightarrow 0$, 
which is incompatible with our assumptions. Hence $H$ and 
$\phi$ go to infinity.

This result explains the examples available in the literature, in which an 
unbounded potential with $dV/d\phi>0$ never 
produces a late time de Sitter solution but rather gives a universe that 
expands faster than exponentially. Sami and 
Toporensky \cite{SamiToporensky03} 
and Guo {\em et al.} \cite{GuoPiaoZhang} consider 
the simple  potential 
$V(\phi)=m^2\phi^2/2$. Let us consider the region $\phi>0$ where 
$dV/d\phi$ is positive -- there $\dot{\phi}$ and $\dot{H}$ approach 
constant values at 
late times \cite{SamiToporensky03},
\be  \label{massivephiattractor}
\dot{\phi} \approx  \sqrt{ \frac{2}{3\kappa }} \, m \;, \;\;\;\;\;
\dot{H} \approx \frac{\kappa \, m}{\sqrt{3} } \, \dot{\phi} 
\ee
as $t\rightarrow +\infty$, and $H, \phi \propto t$ asymptotically. The 
motion becomes potential--dominated at large times and the ratio of 
kinetic to potential energy $
\frac{ -\dot{\phi}^2}{2V} \approx \frac{-\dot{\phi}^2}{m^2 \dot{\phi}^2 
t^2} \rightarrow 0 $ 
so that $V >> \dot{\phi}^2/2$ asymptotically. Both $H$ and $\phi$ diverge 
and the scale factor $a(t)=a_0 \, \mbox{e}^{\alpha t^2/2} $ diverges 
faster than exponentially. However, it takes an 
infinite time for $\phi$ and $H$ to reach infinity and there is no Big 
Rip. 
The effective  equation of state parameter 
\be
w \equiv \frac{P}{\rho}=\frac{ 
\dot{\phi}^2+m^2\phi^2}{\dot{\phi}^2-m^2\phi^2} \rightarrow -1
\ee
as $t \rightarrow + \infty$. The asymptotic equation of state approximates 
$P=-\rho$, which is the exact equation of state for de Sitter space, 
but the solution is 
not de Sitter space nor does it approach it. Asymptotically, 
$\ddot{\phi}\rightarrow 0 $ and the friction and force terms balance each 
other, $ 3H\dot{\phi} \simeq dV/d\phi$. The appendix  is devoted to this 
example and shows that (\ref{massivephiattractor}) is an 
attractor.

Sami and Toporensky \cite{SamiToporensky03} also consider power-law 
potentials $V( \phi) =V_0 
\, \phi^{\alpha}$ (with $ \alpha>0$) and they find a ``slow--climb'' 
regime characterized by
\be
\dot{\phi} \simeq \frac{V'}{3H} \;, \;\;\;\;\; H^2 \simeq 
\frac{\kappa V}{3} \;, \;\;\;\;\; \frac{\dot{\phi}^2}{2V} \simeq 
\frac{1}{6\kappa} \left( \frac{V'}{V} \right)^2 
\ee
if $\alpha < 4$. For $\alpha=4$ they find exponential growth of both 
$\phi$ 
and $H$, while 
if the power-law potential is steeper ($\alpha 
>4$), they find a Big Rip solution with $\phi(t) \approx \left( t_0-t 
\right)^{\frac{2}{4-\alpha}}$. This result is confirmed by Guo {\em 
et al.} \cite{GuoPiaoZhang} and agrees with the qualitative 
argument on the steepness of the potential $V( \phi)$.\\

Hao and Li \cite{HaoLihepth0303093} consider the exponential potential 
$V(\phi)=V_0 \exp  \left( c\phi \right)$ (with $c>0$), which is 
unbounded and has $dV/d\phi>0$ everywhere. They find an 
attractor with equation of state parameter $w<-1$ which makes the Big Rip 
unavoidable. In \cite{HaoLiastroph0309746} the same authors consider 
again the exponential potential $
V(\phi)=V_0 \exp \left[ \sqrt{3} \, \kappa A \left( \phi -\phi_0 \right) 
\right] $ 
with $V_0, A>0$  finding a Big Rip attractor in parametric 
form $\left( a(\phi), t(\phi) \right)$. By eliminating the parameter 
$\phi$ one finds the scale factor
\be
a(t)=\frac{a_0}{\left( \frac{ \sqrt{6V_0} \kappa A^2}{2\sqrt{A^2+2} } 
\right)^{\frac{2}{3A^2} }} \left[
\frac{ 2\sqrt{ A^2+2} }{ \sqrt{6V_0}\kappa A^2} +t_0-t 
\right]^{-\frac{2}{3A^2}} \;.
\ee
In simple terms,  the effective 
equation of state parameter is constant, $w=-(1+A^2)$, and  the scale 
factor can be written as $a\propto \left(t_* -t\right)^{\frac{2}{3(w+1)} 
}$. It is well known that a spatially flat 
FLRW universe filled with a fluid with 
constant equation of state $P=w\rho$ has scale factor $a \propto 
t^{\frac{2}{3(1+w)} }$ \cite{Weinberg}.

It seems intuitive that a potential $V(\phi)$ unbounded 
and steeper than the exponential potential will always cause a sudden 
future 
singularity. This can be shown for  extremely steep potentials: {\em if 
$V(\phi)$ has a vertical asymptote the universe 
evolves to a future singularity in a finite time.}

In fact, since $V(\phi) \rightarrow +\infty$ as $\phi 
\rightarrow \phi_c$, it is $dV/d\phi>0$. $V$ 
is unbounded and strictly increasing and, due to our previous result, 
$\dot{\phi}$ cannot tend to zero. The only two possibilities are that 
$\dot{\phi}$ tends to a finite limit $\dot{\phi}_c $ or that 
$\dot{\phi}$ 
diverges. In both cases $\phi$ reaches the critical value $\phi_c$ in a 
finite time and $V$ and $dV/d\phi$ diverge. 
$\dot{\phi}$ cannot have a finite limit $\dot{\phi}_c$: in fact, if this 
happens, then $\dot{H}=\kappa\dot{\phi}^2/2 \rightarrow \dot{H}_c\equiv 
\kappa\dot{\phi}^2_c/2 $ and $\ddot{\phi}\rightarrow 0$. Then the 
modified Klein--Gordon equation (\ref{6}) yields $3H\dot{\phi} \approx 
dV/d\phi  \rightarrow +\infty$ asymptotically and $H$ must diverge in a 
finite time, which is in contradiction with $\dot{H}\simeq $const. Hence it must be $\dot{\phi}\rightarrow +\infty$ and  
$\dot{H} =\kappa\dot{\phi}^2/2 \rightarrow+\infty$. It may happen that 
both $H$ and $\phi$ diverge or that they stay finite with their 
derivatives diverging at a sudden future singularity.
Examples of this kind of singularities have been found in other models of 
dark energy based on the tachyonic field \cite{Gorinietal03}, the brane 
world  \cite{ShtanovSahni02}, and cosmology with the Gauss--Bonnet term
\cite{ToporenskySujikawa}.

Out last result is consistent with the numerical examples of   
\cite{HaoLihepth0303093,SamiToporensky03}, who find a sudden future 
singularity for potentials with a vertical asymptote.

\section{Discussion and conclusions}
\setcounter{equation}{0}

The dynamics of phantom cosmology are studied without 
assuming specific  potentials $V( 
\phi)$, but assuming that the phantom dark energy has already 
come to dominate the cosmic dynamics. 
In the literature there are also scenarios in which the late time state of 
the universe is dominated again by cold (dark and ordinary) matter with 
zero pressure and the phantom energy decays \cite{HaoLiastroph0309746}:
such scenarios are  {\em a priori} excluded from our analysis.

A clear and unified picture of the dynamics is obtained:  the phase 
space is a two--dimensional  surface in the $\left( H, \phi, \dot{\phi} 
\right)$ space and is composed of two sheets. All the equilibrium points, 
which exist if and only if  $dV/d\phi$ has  zeros, are de 
Sitter 
spaces located on the curve where the two sheets touch each other -- they 
are  given by eqs.~(\ref{12}). 
$H$ is always increasing except when $\dot{\phi}$ vanishes, and there are 
no periodic orbits. In the presence of a bounded potential $V( \phi)$ the 
universe has a 
late time de Sitter attractor whose attraction basin is the entire phase 
space. If $V(\phi)$ is unbounded 
there cannot be  a de Sitter attractor and the late time cosmic expansion 
is either super--exponential with infinite lifetime or else the 
universe ends in a Big Rip or other sudden future singularity. 
Unfortunately, in the general situation,  we cannot provide  a sharp 
criterion on the potential to discriminate between these two 
possibilities. Many results 
for specific potentials that are found in the 
literature are recovered and predictions are made for new scenarios 
featuring potentials with the shapes discussed.

It is worth remembering that there are serious doubts on whether phantom 
fields can be stable \cite{Carrolletal,Clineetal,BuniyHsu} and that the 
classical 
equations of motion considered have to be 
superseded by a semiclassical treatment near the Big Rip singularity. 
Indeed there is some evidence that semiclassical backreaction may avoid 
the Big Rip \cite{escape}.

Finally, there are compelling arguments for a field with canonical 
kinetic energy to couple 
nonminimally to the Ricci curvature \cite{FaraoniPRD96,mybook}. In 
particular, conformal coupling  is associated with an infrared fixed point 
of the renormalization group \cite{rengroup} and avoids causal 
pathologies \cite{SonegoFaraoni}. It is not clear whether the same 
arguments 
apply to a phantom field and force it to couple nonminimally 
(a nonminimally coupled phantom is considered in 
\cite{Szdlowskietal04,oscillators}). The slow--climb regime 
for a phantom would be altered by the inclusion of non--minimal coupling  
terms in 
eqs.~(\ref{4})--(\ref{6}), similarly to what happens in slow-roll 
inflation \cite{AbbottFutamaseMaeda89}. The dynamics of a nonminimally 
coupled phantom will be studied elsewhere.

\section*{Acknowledgments}

The author thanks Werner Israel for pointing out Ref.~\cite{Belinskyetal}. 
This work was supported by the Natural Sciences and Engineering Research 
Council of Canada ({\em NSERC}).

\section*{Appendix: massive phantom field}
\def\theequation{A.\arabic{equation}}\setcounter{equation}{0}

Here we consider a phantom field with potential  
$V(\phi)=m^2 \phi^2 /2 $ as an example.  The phase space is the 
two--dimensional surface $ \frac{ 6}{\kappa} H^2+ 
\dot{\phi}^2 = m^2 \phi^2 $.  By using the dimensionless variables
\be
x \equiv  \sqrt{\kappa} \, H \;, \;\;\;\;\;
y  \equiv  \frac{\kappa m}{ \sqrt{6}}  \, \phi \;, \;\;\;\;\;
z  \equiv  \frac{ \kappa}{ \sqrt{6}} \, \dot{\phi}  \;, 
\ee
the equation of the surface can be written as $
y=\pm \sqrt{ x^2 + z^2} $, which describes a cone with axis along the 
$y$--axis and vertex in 
the origin. The upper sheet is described by $ y=+\sqrt{ x^2+z^2}$ and the 
lower sheet by 
$ y= - \sqrt{ x^2+z^2}$; the two sheets join on the plane $z=0$ along the 
lines $y=\pm x$. 
Although we restrict our analysis to spatially flat ($K=0$) universes, it 
can be shown that the cone separates trajectories belonging to an open 
($K=-1$) universe lying inside the cone from orbits belonging to a closed 
($K=+1$) universe lying outside the cone (cf. \cite{Belinskyetal}).

The only equilibrium point satisfying eqs.~(\ref{12}) is 
the Minkowski space  
$\left( x, y,z \right)=\left( 0,0,0 \right)$. 
Our results imply that $H$ and $\phi$ go to 
infinity for any solution except the equilibrium point, and hence the 
equilibrium point is unstable.  Since $V''(0)=m^2 >0$ the 
fixed point (which coincides with the global minimum of the potential) is 
unstable -- see the discussion of Sec.~III.

This three-dimensional picture should be compared with fig.~3 of 
\cite{GuoPiaoZhang} showing a projection of the phase space (a 
two-dimensional cone) and of 
some orbits onto the $\left( \phi, \dot{\phi} \right) $ plane, 
corresponding to our $\left( y,z \right)$ plane. A 
similar conical geometry of the phase space structure appears if one 
considers inflation realized by a massive scalar with the ``right'' sign 
of the kinetic energy density \cite{Belinskyetal}.

The trajectories of the solutions exhibit  a late time attractor found 
analytically in \cite{SamiToporensky03}
(eq.~(\ref{massivephiattractor})). The 
authors  of \cite{SamiToporensky03,GuoPiaoZhang} show numerically 
the convergence of many solutions to this asymptotic solution but do not 
prove analytically its attractor nature. The attractor can be 
obtained by requiring   
an asymptotic linear relationship between $H$ and $\phi$, as suggested by 
numerical integrations. By setting $H=\alpha \phi$ one obtains, using 
eq.~(\ref{15}), $\dot{\phi} = 2\alpha/\kappa$. Eq.~(\ref{4}) yields 
$
H \approx \pm \sqrt{ \frac{2}{3\kappa} } \, \alpha\, m \, t $,
while eq.~(\ref{5}) is identically satisfied and eq.~(\ref{6}{) gives 
$\alpha=m \sqrt{ \kappa/6} $.  Since $\dot{H} \geq 0$ the negative sign 
is rejected. The asymptotic solution found in this 
way coincides with  the one of \cite{SamiToporensky03}
\be
\left( H_* , \phi_* \right) =\left(  \frac{m^2}{3} \, t ,   
 \sqrt{ \frac{2}{3\kappa} } \, m \, t \right)  \;. 
\ee
 The stability with 
respect to linear homogeneous perturbations is assessed by considering the 
perturbed Hubble parameter and scalar field
$ H=H_* +\delta H $, $ \phi=\phi_* +\delta \phi $, 
and inserting these expressions into the evolution equations 
(\ref{4})--(\ref{6}). It is convenient to consider the contrasts
$
\Delta_1 \equiv \delta H/H_* $ and $\Delta_2 \equiv \delta \phi/\phi_* $. 
One finds $
\Delta_1=\sqrt{ \frac{3\kappa}{2}} \, \frac{1}{mt} \left( \delta \phi 
-\frac{\delta \dot{\phi}}{m^2 t} \right) $ 
and eq.~(\ref{6}) yields the evolution equation for $\Delta_2$
\be
\ddot{\Delta_2} + \left( \frac{2}{t}-m + m^2 t \right) \dot{\Delta_2} 
+ \left(  m^2-\frac{1}{t^2} \right) \Delta_2 =0
\ee
with asymptotic solution $ \Delta_2  \simeq$const. and $
\Delta_1 = \Delta_2 \left( 1-\frac{1}{m^2t^2} \right) -
\frac{ \dot{\Delta_2}}{m} \simeq \mbox{constant} $. 
The  solution $\left( H_*, \phi_* \right) $ is stable and 
is an attractor, in agreement with the numerical results.


{\small \begin{thebibliography}{99}

\bibitem{SN} Riess A G {\em et al.} 19998  {\em Astron. J.}
{\bf 116} 1009;  Perlmutter S {\em et al.} 1998 {\em
Nature} {\bf 391} 51; Riess A G  {\em et al.} 1999 {\em Astron.
J.} {\bf 118} 2668;  Perlmutter S {\em et al.} 1999 {\em
Astrophys. J.} {\bf 517} 565; Riess A G {\em et al.} 2000 {\em
Astrophys. J.} {\bf 536} 62;  Riess A G {\em et al.} 2001 {\em
Astrophys. J.} {\bf 560} 49

\bibitem{WMAP}  Bennett C {\em et al.} 2003 {\em Astrophys. J. (Suppl.)} 
{\bf 148} 1;  
Hinshaw G {\em et al.} 2003 {\em Astrophys. J. (Suppl.)} {\bf 148} 135; 
Kogut A {\em et al.} 2003 {\em Astrophys. J. (Suppl.)} 
{\bf 148} 161; 
Bridle S {\em et al.} 2003 {\em Science} {\bf 239} 1532; 
Spergel D N {\em et al.} 2003 {\em Astrophys. J. (Suppl.)} 
{\bf 148} 175; Page L {\em et al.} 2003 
{\em Astrophys. J. (Suppl.)} {\bf 148} 233

\bibitem{LSS} Allen S W {\em et al.} 2004, {\em Mon. Not. R. Astr. Soc.} 
{\bf 353} 4571

\bibitem{modifiedgravity} 
Capozziello S 2002 {\em Int. J. Mod. Phys. D} {\bf 11} 483;  
Capozziello S, Carloni S and  Troisi M 2003, 
{\em Int. J. Mod. Phys. D} {\bf 12} 1969; 
Carroll S M , Duvvuri V, 
Trodden M and Turner M S 2004 {\em Phys. Rev. D} {\bf 70} 043528;
Nojiri S and  Odintsov S D 2003 {\em Phys. Rev. D} {\bf 68} 123512; 
2004 {\em Phys. Lett. B} {\bf 595} 1; 2004 {\em Mod. Phys. Lett. A} {\bf 
19} 627; 
2004 {\em Gen. Rel. Grav.} {\bf 36} 1765; 
Abdalla M C B, Nojiri S and Odintsov S D, hep-th/0409172;
Elizalde E {\em et al.} 2003 {\em Phys. Lett. B} {\bf 574} 1; 
Elizalde E,  Nojiri S and  Odintsov S D 2004 {\em Phys. Rev. D} {\bf 70} 
043539;  
Chiba T 2003, {\em Phys. Lett. B} {\bf 575} 1;
Cognola G and Zerbini S, gr-qc/0407103;
Soussa M E and Woodard R P 2004 {\em Gen. Rel. Grav.} {\bf 36} 855
 
\bibitem{Palatini} 
Vollick D N 2003 {\em Phys. Rev. D} {\bf 68} 063510; 
2004 {\em Class. Quant. Grav.} {\bf 21} 3813;
2004 {\em Phys. Rev. D} {\bf 69} 064030; 
2004, {\em Class. Quant. Grav.} {\bf 21} 3813;
Flanagan E E 2004 {\em Phys. Rev. Lett.} {\bf 92} 071101;
2004 {\em Class. Quant. Grav.} {\bf 21} 3817;
Meng X H and Wang P 2004 {\em Class. Quant. Grav.} {\bf 20} 4949; 
2004 {\em Class. Quant. Grav.} {\bf 21} 951;
2004 {\em Gen. Rel. Grav.} {\bf 36} 1947;
hep-th/0310038;
2004 {\em Phys. Lett. B} {\bf 584} 1;
astro-ph/0308284;
Wang P,  Kremer G M , Alves D S M and  Meng X--H, gr-qc/0408058;
Kremer G M and  Alves D S M 2004 {\em Phys. Rev. D} {\bf 70} 023503;
Olmo G J  and  Komp W, gr-qc/0403092; 
Allemandi G, Capone M , 
Capozziello S and  Francaviglia M, hep-th/0409198


\bibitem{mginstabilities} Chiba T 2003 {\em Phys. Lett. B} {\bf 575} 1; 
Dolgov A D and Kawasaki M 2003 {\em Phys. Lett. B} {\bf 573} 1; 
Soussa M E and Woodard R P 2004 {\em Gen. Rel, Grav.} {\bf 36} 855;
Nunez A and Salganik S 2005 {\em Phys. Lett. B} {\bf 608} 189


\bibitem{mglimit} Lue A, Scoccimarro R and Starkman G D 2004, 
{\em Phys. Rev. D} {\bf 69} 124015;
Dick R 2004, {\em Gen. Rel. Grav.} {\bf 36} 217;
Dominguez A E and  Barraco D E 2004, {\em Phys. Rev. D} {\bf 70}, 043505


\bibitem{Caldwell}  Caldwell R R 2002 {\em Phys. Lett. B} {\bf 545} 23


\bibitem{Melchiorrietal03} Melchiorri A, Mersini L,  Odman C J and  
Trodden M 2003 {\em Phys. Rev. D} {\bf 68} 043509


\bibitem{superq} Faraoni V 2002 {\em Int. J. Mod. Phys. D} {\bf 11} 471


\bibitem{mybook} Faraoni V 2004, {\em Cosmology in Scalar--Tensor Gravity} 
(Dordrecht: Kluwer Academic)

\bibitem{Caldwelletal} Caldwell R R , Kamionkowski M and
Weinberg N N 2003 {\em Phys. Rev. Lett.} {\bf 91} 071301


\bibitem{Carrolletal} Carroll S M ,  Hoffman M and Trodden M 2004 {\em 
Phys. Rev. D} {\bf 68} 023509


\bibitem{Barrow04}  Barrow J D 2004 {\em Class. Quant. Grav.} {\bf 21}  
L79;   Lake K 2004 {\em Class. Quant. Grav.} {\bf 21} L129;
Frampton P H,  astro-ph/0407353;
Chimento L P and Lazkoz R 2004 {\em Mod. Phys. Lett. A} {\bf 19} 2479


\bibitem{FaraoniPRD2003} Faraoni V 2003 {\em Phys. Rev. D} {\bf 68} 
063508; 2000 {\em Phys. Rev. D} {\bf 62} 023504


\bibitem{FujiiMaeda}  Fujii Y and Maeda K 2003 {\em The Scalar--Tensor 
Theory of Gravity} (Cambridge: Cambridge University Press)


\bibitem{CarvalhoSaa}  Carvalho F C and Saa A 2004 {\em Phys. Rev. D} 
{\bf 70} 087302


\bibitem{MersiniBasteroKanti} Mersini L,  Bastero--Gil M and  Kanti P 2001 
{\em Phys. Rev. D} {\bf 64} 043508

\bibitem{BasteroFramptonMersini} Bastero--Gil M , Frampton P H and  
Mersini L 2002 {\em Phys. Rev. D} {\bf 65} 106002  

\bibitem{Frampton03} Frampton P H 2003 {\em Phys. Lett. B} {\bf 555}  
139  

\bibitem{Nilles84} Nilles H P 1984 {\em Phys. Rep.} {\bf 110} 1

\bibitem{Barrow88} Barrow J D 1988 {\em Nucl. Phys. B} {\bf 310} 743 

\bibitem{Pollock88} Pollock M D 1988 {\em Phys. Lett. B} {\bf 215}
635 

\bibitem{Sergei03} Nojiri S and Odintsov S D 2003, {\em Phys. Lett. B} 
{\bf 562} 147

\bibitem{Clineetal} Cline J M,  Jeon S and  Moore G D 2004 {\em Phys. 
Rev. D} {\bf 70} 043543 

\bibitem{BuniyHsu}  Hsu S D H , Jenkins A and Wise M B 2004 {\em Phys. 
Lett. B} {\bf 597} 270;
Buniy R V and Hsu S D H, preprint 
hep-th/0502203

\bibitem{VisserKarDadhich03} Visser M,  Kar S and  Dadhich N 2003 {\em 
Phys. Rev. Lett.} {\bf 90}  201102  

\bibitem{BarceloVisser} Barcelo C and Visser M 2000 {\em Class. Quant. 
Grav.} {\bf 17} 3843; 
Bellucci S and  Faraoni V 2002 {\em Nucl. Phys. B} {\bf 640}, 453;
Hiscock W A 1990 {\em Class. Quant. Grav.} {\bf 
7} L35;
Deser S 1984 {\em Phys. Lett. B} {\bf 134}, 419 

\bibitem{oscillators} Faraoni V 2004 {\em Phys. Rev. D} {\bf 69} 123520 

\bibitem{phantommodels} Fang W {\em et al.}, hep-th/0409080; 
Gonzalez--Diaz P F and Jimenez--Madrid J A 2004 {\em Phys. Lett. B} {\bf 
596} 16;
Brown M G, Freese K and Kinney W H, astro-ph/0405353;
Elizalde E,  Nojiri S and Odintsov SD 2004 {\em Phys. Rev. D} {\bf 70} 
043539;
Hao J--G and  Li X--Z 2005 {\em Phys. Lett. B} {\bf 606} 7;
Aguirregabiria J M, Chimento L P and Lazkoz R 2004 {\em Phys. Rev. D} 
{\bf 70} 023509;
Piao Y--S  and  Zhang Y--Z 2004 {\em Phys. Rev. D} {\bf 70} 063513;
Gonzalez--Diaz P F 2003 {\em Phys. Rev. D} {\bf 68} 021303(R);
2004 {\em Phys. Rev. D} {\bf 69} 063522;
2004 {\em Phys. Lett. B} {\bf 586} 1;
Lu H Q, hep-th/0312082;
Johri V B 2004 {\em Phys. Rev. D} {\bf 70} 041303;
Stefancic H 2004 {\em Phys. Lett. B} {\bf 586} 5;
Liu D--J and  Li X--Z 2003 {\em Phys. Rev. D} {\bf 68} 067301;
Hao J--G  and  Li X--Z 2003 {\em Phys. Rev. 
D} {\bf 68} 043501;
Li X--Z and  Hao J--G 2004 {\em Phys. Rev. 
D} {\bf 69}  107303;
Dabrowski M P, Stachowiak T and Szdlowski M 2003 {\em Phys. Rev. D} {\bf 
68} 103519;
Babichev V,  Dokuchaev V and  Eroshenko Yu 2004 {\em Phys. Rev. Lett.} 
{\bf 93} 021102 

\bibitem{STphasespace}  Faraoni V 2005 {\em Ann. Phys. (NY)} {\bf 317} 366

\bibitem{Wald} Wald R M 1984 {\em General Relativity} (Chicago: Univ. of 
Chicago Press)

\bibitem{SinghSamiDadhich03} Singh P, Sami M and Dadhich N 2003 {\em 
Phys. Rev. D} {\bf 68}  023522  

\bibitem{SamiToporensky03} Sami M and  Toporensky A 2004 {\em Mod. Phys. 
Lett. A} {\bf 19} 1509 

\bibitem{Amendolaetal} Amendola L, Litterio M and Occhionero F 1990 {\em 
Int. J. Mod. Phys. A} {\bf 5} 3861  

\bibitem{Foster} Foster S 1998 {\em Class. Quant. Grav.} {\bf 15} 3485  

\bibitem{Gunzigetal}  Gunzig E {\em et al.} 2001 {\em Phys. Rev. D} {\bf 
63} 067301; 2000 {\em Class. Quant. Grav,} {\bf 17} 1783; 
2000 {\em Int. J. Theor. Phys.} {\bf 39} 1901; 
Saa A {\em et al.} 2001 {\em Int. J. Theor. Phys.} {\bf 40}  2295; 
Rocha--Filho T M {\em et al.} 2000 {\em Int. J. Theor. Phys.} {\bf 39} 
1933  


\bibitem{HaoLi03PRD2}  Hao J--G and  Li X--Z 2003 {\em Phys. Rev. D} 
{\bf 67} 107303 

\bibitem{HaoLiastroph0309746} Hao  J--G and  Li X--Z 2004 {\em Phys. Rev. 
D} {\bf 70} 043529 

\bibitem{deSitter} Faraoni V 2004 {\em Phys. Rev. D} {\bf 70} 044037 

\bibitem{GuoPiaoZhang}  Guo Z--K,  Piao Y--S and Zhang Y--Z 2004
{\em Phys. Lett. B} {\bf 594} 247

\bibitem{HaoLihepth0303093} Hao J--G and  Li X--Z 2004 {\em Phys. Rev. D} 
{\bf 69} 107303

\bibitem{Weinberg} Weinberg S 1972 {\em Gravitation and Cosmology} (New 
York: Wiley)

\bibitem{Gorinietal03}  Gorini V,  Kamenshchik A Yu,  Moschella U and  
Pasquier V 2004 {\em Phys. Rev. D} {\bf 69} 123512 

\bibitem{ShtanovSahni02}  Shtanov Yu and Sahni V 2002 {\em Class. Quant. 
Grav.} {\bf 19} L101 

\bibitem{ToporenskySujikawa} Toporensky A and  Tsujikawa S 2002 {\em Phys. 
Rev. D} {\bf 65} 123509 

\bibitem{Szdlowskietal04}  Szdlowski M,  Czaja W and  Krawiec A, 
astro-ph/0401293

\bibitem{HaoLi03PRD3} Hao J--G and  Li X--Z 2003 {\em Phys. Rev. D} 
{\bf 68} 083514; 2004 {\em Phys. Rev. D} {\bf 70} 043529 

\bibitem{escape} Nojiri S and  Odintsov S D 2004 {\em Phys. Lett. B} {\bf 
595} 1;  Elizalde E,  Nojiri S, 
Odintsov S D and  Wang P, hep-th/0502082; 
Nojiri S,  Odintsov S D and  Tsujikawa S 2005 {\em Phys. Rev. D} {\bf 71} 
063004; 
 Elizalde E, Nojiri S and  Odintsov S D 2004 {\em 
Phys. Rev. D} {\bf 70} 043539;
Linder E V  2004 {\em Phys. Rev. D} {\bf 70} 023511;
Wu P and Yu H, astro-ph/0407424


\bibitem{FaraoniPRD96} Faraoni V 1996 {\em Phys. Rev. D} {\bf 53} 6813; 
Faraoni V,  Gunzig E and  Nardone P 1999, {\em Fundam. Cosm. Phys.} 
{\bf 20} 121 

\bibitem{rengroup} 
Buchbinder I L and  Odintsov S D 1983 {\em Sov. J. Nucl. Phys.} 
{\bf 40} 848; 1985 {\em Lett. Nuovo
Cimento} {\bf 42} 379; 
Buchbinder I L 1986 {\em Fortschr. Phys.} {\bf 34} 605; 
Buchbinder I L, Odintsov S D and Shapiro I L 1989, {\em Riv. Nuovo 
Cimento} {\bf 12} 1;
Odintsov S D 1991 {\em Fortschr. Phys.} {\bf 39} 621; 
Muta T S and Odintsov S D 1991 {\em Mod.
Phys. Lett. A} {\bf 6} 3641;
Elizalde E and Odintsov S D 1994 {\em Phys.
Lett. B} {\bf 333} 331 

\bibitem{SonegoFaraoni} Sonego S and  Faraoni V 1993 {\em Class. Quant. 
Grav.} {\bf 10} 1185 

\bibitem{AbbottFutamaseMaeda89} Abbott L F 1981 {\em Nucl. Phys. B} 
{\bf 185} 233; 
Futamase T and  Maeda K I 1989 {\em Phys. Rev. D} 
{\bf 39} 399 

\bibitem{Belinskyetal} Belinsky V A,  Grishchuk L P,  Khalatnikov I M 
and  Zeldovich Ya B 1985 {\em Phys. Lett. B} {\bf 155} 232 


\end{thebibliography}}
\end{document}